\documentclass[aps,prd,preprint,tightenlines,%superscriptaddress,
 nofootinbib]{revtex4}

\let\k=\kappa

\newcommand{\be}{\begin{equation}}
\newcommand{\ee}{\end{equation}}
\newcommand{\bea}{\begin{eqnarray}}
\newcommand{\eea}{\end{eqnarray}}

\newcommand{\PRE}[1]{{#1}}   % Use if preprint style
\usepackage{bm}
\usepackage{epsfig}
\begin{document}

 \preprint{UCI-TR-2003-55}

\title{
\PRE{\vspace*{1.5in}} Newtonian Gravity in Theories with Inverse
Powers of R\PRE{\vspace*{0.3in}} }

\author{Arvind Rajaraman}
\affiliation{Department of Physics and Astronomy, University of
California, Irvine, CA 92697, USA \PRE{\vspace*{.3in}} }

%\date{July 2003}

\begin{abstract}
\PRE{\vspace*{.1in}}We discuss  theories with inverse powers of
the Ricci scalar.
 We find the exact metric produced by point masses in these theories, and
 show that the gravitational force between static objects is consistent with
 experiments.
\end{abstract}

\maketitle

\section{Introduction}
  Recent astrophysical measurements seem to indicate that we live in
  an universe dominated by dark energy. The nature of this energy
  is
  still unknown. It might be a cosmological constant, or
  quintessence.

  It is also possible that these new observations actually
  indicate
  that gravity needs to be modified on very long distance scales.
 Several models that produce such modifications exist in the
 literature.
  We will be particularly concerned with the model proposed
  in \cite{Carroll:2003wy}, where the gravitational action was modified by the
  addition of inverse powers of the Ricci scalar.

  Explicitly, these authors considered an action of the form
  \bea \label{1RLag}
  S=\int d^4x \sqrt{-g}\left(R-{\mu^4\over R}\right)
  \eea
  (We  shall sometimes refer to this as the 1/R theory).
  This action is clearly singular at $R=0$, and accordingly,
  Minkowski space
  is not a solution of this model.
  This model does possess solutions which are de Sitter or
  Anti-de Sitter. Therefore, the effects of dark energy
  can be reproduced by this modification of
  gravity\footnote{For previous work on this subject, see
  \cite{Vollick:2003aw,
 Dick:2003dw,Dolgov:2003px,Chiba:2003ir,Flanagan:2003iw,
 Nojiri:2003ft,Meng:2003uv,Flanagan:2003rb,Nojiri:2003rz,Nojiri:2003wx}.}.

  It would seem that if $\mu$ is sufficiently small, then the
  effects of the new term should be negligible except on
  cosmological length scales $r\sim \mu^{-1}$. However, this
  intuition has been challenged by a number of authors \cite{Vollick:2003aw,
 Dick:2003dw,Dolgov:2003px,Chiba:2003ir,Flanagan:2003iw}. The claim
  in these papers is that, in fact, the above theory is equivalent
  to a scalar-tensor theory, where the scalar field couples
  gravitationally and has an extremely tiny mass $m\sim \mu$.
  Such a theory is ruled out on many grounds, for instance
  solar system measurements.

  Here we shall look at some solutions of this theory, in
  a hope to better understand the issues. In particular, we will find
  the {\it exact} solution for the field of a point mass in this
  theory. Remarkably, the solution is identical to the solution
  for a point mass in general relativity with a cosmological constant.
  Given the solution, it is straightforward to
  find the force between point particles and show that it is
  consistent with  experiments. We discuss the possible significance of
  this solution.

\section{ Gravitational Source of point particles in 1/R theory}

 Here we  find the exact
gravitational solution for a point mass in the 1/R theory. We
shall use the solution to find the forces between point masses and
show that they are consistent with experiment.

 The action\bea
  S={1\over 2\k^2}\int d^4x \sqrt{-g}\left(R-{\mu^4\over R}\right)+S_m
  \eea
($S_m$ is the matter action) has the equation of motion
\bea\label{eom} \left(1+{\mu^4\over R^2}\right) R_{\mu\nu}-{1\over
2}\left(1-{\mu^4\over R^2}\right) Rg_{\mu\nu}
+\mu^4\left[g_{\mu\nu}D^2-D_{(\mu} D_{\nu)}\right]R^{-2}={\k^2
T_{\mu\nu}} \eea

Away from the localized sources, the right hand side can be set to
zero.

 We will start by looking for solutions of the form
 $
 R_{\mu\nu}=\pm \Lambda g_{\mu\nu}
$.
  We then find
  \bea
  \Lambda={\sqrt{3}\over 4}\mu^2
  \eea

So we find that away from source terms, any solution satisfying
$R_{\mu\nu}=\pm {\sqrt{3}\over 4}\mu^2 g_{\mu\nu}$ also satisfies
the equation of motion (\ref{eom}). That is, any solution of
Einstein's equations with a positive {\it or} negative
cosmological constant is a solution of the 1/R theory. This means
that we can immediately write down a large class of solutions to
equation (\ref{eom}), just by looking at solutions of
 $R_{\mu\nu}=\pm \Lambda g_{\mu\nu}$.

The simplest such solutions are:

%de Sitter and anti de Sitter
%space.
  a) de Sitter space

\bea ds^2=-\left(1-{\Lambda r^2\over 3}\right) dt^2
+\left(1-{\Lambda r^2\over 3}\right)^{-1} dr^2
+r^2(d\theta^2+\sin^2\theta d\phi^2)\eea

and  b) Anti-de Sitter space\bea ds^2=-\left(1+{\Lambda r^2\over
3}\right) dt^2 +\left(1+{\Lambda r^2\over 3}\right)^{-1} dr^2
+r^2(d\theta^2+\sin^2\theta d\phi^2)\eea

 They represent the two
vacuum solutions of the 1/R theory.

The remarkable feature here is that we could write down these
solutions just by looking at previously known solutions of general
relativity (GR) with a cosmological constant. This may clearly be
extended to localized sources. Away from the sources, we solve
$R_{\mu\nu}=\pm \Lambda g_{\mu\nu}$  with several localized
sources (the sign is fixed by the boundary conditions.) This is
now a problem in standard GR, and can be addressed using standard
methods. The solution of the GR problem will produce a solution of
(\ref{eom}) with the same sources. This applies to diffuse and
time-dependent sources as well.

For a point source, the procedure outlined above yields  the
Schwarzschild-de-Sitter  and Schwarzschild-Anti-de Sitter
solutions which are of the form

  c) Schwarzschild-de Sitter
 (black holes in de Sitter space)

\bea \label{deSch} ds^2=-\left(1-{2M\over r}-{\Lambda r^2\over
3}\right) dt^2 +\left(1-{2M\over r}-{\Lambda r^2\over
3}\right)^{-1} dr^2 +r^2(d\theta^2+\sin^2\theta d\phi^2)\eea

 and d) Schwarzschild- anti de Sitter (black holes in anti de Sitter
space)

 \bea ds^2=-\left(1-{2M\over
r}+{\Lambda r^2\over 3}\right) dt^2 +\left(1-{2M\over r}+{\Lambda
r^2\over 3}\right)^{-1} dr^2 +r^2(d\theta^2+\sin^2\theta
d\phi^2)\eea

Outside the horizon, these satisfy the equation  $R_{\mu\nu}=\pm
\Lambda g_{\mu\nu}$, and therefore also satisfy the equation
(\ref{eom}). These solutions represent the field produced by a
point mass in the 1/R theory. To determine the solution fully, we
still have to choose de-Sitter or anti-de-Sitter boundary
conditions.

\section{Discussion}

Let us return to the original question: the force between two
masses in  the 1/R theory. Since we are comparing with experiments
in the real world, we should presumably take de Sitter boundary
conditions. The field of each mass is then the de
Sitter-Schwarzschild solution (\ref{deSch}). The gravitational
force can then be determined by taking the Newtonian limit.

The important point is that as $\mu$ gets small, the
de-Sitter-Schwarzschild solution reduces to the Schwarzschild
solution, so we are assured that we recover the standard Newtonian
force law in the $\mu\rightarrow 0$ limit. The corrections to
Newtonian gravity will show up at distance scales when $\Lambda
r^2\sim GM/r$, i.e. $r\sim (GM/\Lambda)^{1/3}$. For solar masses,
this scale is roughly $10^{17}$ m, much larger than the solar
system. So the deviations from Newtonian gravity on the scale of
the solar system are negligible.

This appears to show that at least static solutions to these
equations are in agreement with Newtonian gravity. The question is
whether there are other solutions which are not of this form. In
particular, the equations of motion are quartic, and it may be
that further data is needed to specify the field of a point mass.

It might seem that the boundary conditions plus regularity should
suffice to specify the solution for a point mass. However, the
existence of the equivalent scalar-tensor theory with a light
scalar sheds doubt on this speculation. It would be interesting to
explore these issues further. It would be particularly interesting
to see if the requirement of regularity of the metric puts
constraints on the solutions of the scalar-tensor theory.

\section{Acknowledgements}

I would like to thank Fumihiro Takayama, Nima Arkani-Hamed and
Eanna Flanagan for stimulating discussions.

%%%%%%%%%%%%%%%%%%%%%%%%%%%%%%%%%%%%%%%%%%%%%%%%%%%%%%


\begin{thebibliography}{99}
%%%%%%%%%%%%%%%%%%%%%%%%%%%%%%%%%%%%%%%%%%%%%%%%%%%%%%

%\cite{Carroll:2003wy}
\bibitem{Carroll:2003wy}
S.~M.~Carroll, V.~Duvvuri, M.~Trodden and M.~S.~Turner,
%``Is cosmic speed-up due to new gravitational physics?,''
arXiv:astro-ph/0306438.
%%CITATION = ASTRO-PH 0306438;%%

%\cite{Vollick:2003aw}
\bibitem{Vollick:2003aw}
D.~N.~Vollick,
%``Curvature Corrections as the Source of the Cosmological Acceleration,''
Phys.\ Rev.\ D {\bf 68}, 063510 (2003) [arXiv:astro-ph/0306630].
%%CITATION = ASTRO-PH 0306630;%%

%\cite{Dick:2003dw}
\bibitem{Dick:2003dw}
R.~Dick,
%``On the Newtonian Limit in Gravity Models with Inverse Powers of R,''
arXiv:gr-qc/0307052.
%%CITATION = GR-QC 0307052;%%

%\cite{Dolgov:2003px}
\bibitem{Dolgov:2003px}
A.~D.~Dolgov and M.~Kawasaki,
%``Can modified gravity explain accelerated cosmic expansion?,''
Phys.\ Lett.\ B {\bf 573}, 1 (2003) [arXiv:astro-ph/0307285].
%%CITATION = ASTRO-PH 0307285;%%

%\cite{Chiba:2003ir}
\bibitem{Chiba:2003ir}
T.~Chiba,
%``1/R gravity and scalar-tensor gravity,''
arXiv:astro-ph/0307338.
%%CITATION = ASTRO-PH 0307338;%%

%\cite{Flanagan:2003iw}
\bibitem{Flanagan:2003iw}
E.~E.~Flanagan,
%``Higher order gravity theories and scalar tensor theories,''
arXiv:gr-qc/0309015.
%%CITATION = GR-QC 0309015;%%



%\cite{Nojiri:2003ft}
\bibitem{Nojiri:2003ft}
S.~Nojiri and S.~D.~Odintsov,
%``Modified gravity with negative and positive powers of the curvature: Unification of the inflation and of the cosmic acceleration,''
arXiv:hep-th/0307288.
%%CITATION = HEP-TH 0307288;%%

%\cite{Meng:2003uv}
\bibitem{Meng:2003uv}
X.~Meng and P.~Wang,
%``Cosmological Evolution in 1/R-Gravity Theory,''
arXiv:astro-ph/0308031.
%%CITATION = ASTRO-PH 0308031;%%

%\cite{Flanagan:2003rb}
\bibitem{Flanagan:2003rb}
E.~E.~Flanagan,
%``Palatini form of 1/R gravity,''
arXiv:astro-ph/0308111.
%%CITATION = ASTRO-PH 0308111;%%


%\cite{Nojiri:2003rz}
\bibitem{Nojiri:2003rz}
S.~Nojiri and S.~D.~Odintsov,
%``Where new gravitational physics comes from: M-theory,''
Phys.\ Lett.\ B {\bf 576}, 5 (2003) [arXiv:hep-th/0307071].
%%CITATION = HEP-TH 0307071;%%

%\cite{Nojiri:2003wx}
\bibitem{Nojiri:2003wx}
S.~Nojiri and S.~D.~Odintsov,
%``The minimal curvature of the universe in modified gravity and conformal anomaly resolution of the instabilities,''
arXiv:hep-th/0310045.
%%CITATION = HEP-TH 0310045;%%



\end{thebibliography}
\end{document}